\documentclass{jdmdh}
\usepackage{array}
\usepackage{pgfplots}

\title{Tracking the Consumption Junction: Temporal Dependencies between Articles and Advertisements in Dutch Newspapers}
\author[1]{Melvin Wevers}
\author[2]{Jianbo Gao}
\author[3]{Kristoffer L. Nielbo}
\affil[1]{Digital Humanities Lab, KNAW Humanities Cluster, Netherlands} 
\affil[2]{Institute of Complexity Science and Big Data Technology, Guangxi University, China}
\affil[3]{Center for Humanities Computing Aarhus, Aarhus University, Denmark} 

\corrauthor{Kristoffer L. Nielbo}{kln@cas.au.dk}

\begin{document}

\maketitle

\abstract{Historians have regularly debated whether advertisements can be used as a viable source to study the past. Their main concern centered on the question of agency. Were advertisements a reflection of historical events and societal debates, or were ad makers instrumental in shaping society and the ways people interacted with consumer goods? Using techniques from econometrics (Granger causality test) and complexity science (Adaptive Fractal Analysis), this paper analyzes  to what extent advertisements shaped or reflected society. We found evidence that indicate a fundamental difference between the dynamic behavior of word use in articles and advertisements published in a century of Dutch newspapers. Articles exhibit persistent trends that are likely to be reflective of communicative memory. Contrary to this, advertisements have a more irregular behavior characterized by short bursts and fast decay, which, in part, mirrors the dynamic through which advertisers introduced terms into public discourse. On the issue of whether advertisements shaped or reflected society, we found particular product types that seemed to be collectively driven by a causality going from advertisements to articles. Generally, we found support for a complex interaction pattern dubbed the consumption junction. Finally, we discovered noteworthy patterns in terms of causality and long-range dependencies for specific product groups. All in, this study shows how methods from econometrics and complexity science can be applied to humanities data to improve our understanding of complex cultural-historical phenomena such as the role of advertising in society.}


\section{Introduction}

\strut
\vspace{-4ex}

\subsection{Advertisements as a Lens on the Past}

Over the course of the twentieth century, branded consumer goods turned into an integral part of society \cite{cross_all-consuming_2000,de_grazia_irresistible_2005}. This sparked research into consumer goods, as well as research that relies on consumer goods as entry points into broader cultural phenomena. Historians, for instance, have studied how advertisements represented consumerism, gender identities, and the globalization of food cultures \cite{lears_fables_1994,parkin_food_2007,sivulka_soap_2012}. For these scholars, advertisements functioned as a lens on the past. Roland Marchand argues that advertisements provide an insight into the ideals and aspirations of past realities. They show the state of technology, the social functions of products, and provide information on the society in which a product was sold. Furthermore, Marchand poses that advertisements contributed to the shaping of a ``community of discourse.'' He claims that advertisements infused public discourse with a particular type of language \cite{marchand_advertising_1985}.

At the same time, scholars have debated to what extent adverts actually offer a meaningful depiction of the past. In their effort to sell more products, producers, and ad makers amplified or distorted certain social and cultural aspects to make products more appealing to consumers \cite{brandt_engineering_2004}. Erving Goffman points out that ads shaped consumer's lived experience by prescribed certain conceptions of identity \cite{goffman_gender_1985}. Put differently, ad makers used adverts to \textit{shape} society by tweaking the desires of consumers \cite{fox_mirror_1997}. This form of social engineering was rooted in psychological and sociological theories of the day. Especially the tobacco industry was very actively involved in developing new ways of advertising that could entice people to pick up cigarettes \cite{brandt_engineering_2004}.

These theoretical debates raise the question whether we can actually use advertisements to study the past, or are we merely studying a version of the past constructed by ad makers? Theories on the relationship between advertisements and society can be summarized by three positions. The first position contends that advertisements reflected the desires and aspirations of consumers. The second argues that advertisements merely represented the interest of advertisers and the companies that produced the commodities. This approach attributes more agency to the advertisers. The third approach acknowledges that there has existed a more complex ``consumption junction'', in which producers, distributors, consumers, and advertisers collectively negotiated the meaning and success of a consumer product \cite{cowan_proposal_1997}.

This article examines the validity of these three positions in a specific historical context (i.e. twentieth-century Netherlands) and sets out to answer to what extent advertisements reflected or shaped society. We studied this interplay between advertisers and society by analyzing advertisements and articles in newspapers. Newspapers are a well-read rich historical source that contains ``conscious representations of conditions and events'' as well as ``unconscious reflection[s] of the tastes, the interests, the desires, and the spirit of its day \cite[p.48]{smail_deep_2008}.'' As such, newspapers function as a proxy for public discourse \cite{schudson_power_1982,marshall_contesting_1995}. Moreover, the availability of digitized newspapers enables researchers to use computation to explore the archive, locate particular instances of language use, or extract specific linguistic patterns.

This study combines techniques from econometrics and complexity science to examine the dynamics of word use in articles and advertisements. We set out to answer the following three questions. First, did advertisements shape or reflect newspaper discourse? Second, did word use in ads differ in dynamics from articles? And finally, were these characteristics more pronounced for particular product groups? Answering these questions will further our understanding of the function of advertising in society.

\subsection{Beyond Counting Word Frequencies}
With the proliferation of large databases that hold  temporally-dispersed text content, time series plots of word frequencies have become a valuable source of exploration and validation of cultural trends \cite{gao_culturomics_2012,michel_quantitative_2011}. Google has popularized this approach through their web-based n-gram viewer of time-dependent relative word frequencies derived from its digitized book collection.\footnote{ Google Books Ngram Viewer: \url{https://books.google.com/ngrams}} The popularity of the Google Ngram viewer has sparked other digital archives to also develop n-gram viewers based on their collection.\footnote{ Examples of ngram viewers: The British Library: \url{https://www.webarchive.org.uk/ukwa/ngram/}, Danish Royal Library: \url{http://labs.statsbiblioteket.dk/smurf/}, and  National Library of the Netherlands \url{http://kbkranten.politicalmashup.nl}}.

N-gram viewers can help researchers to determine when words appeared and how they evolved. Plots of word frequencies, however, only offer an overview of particular trends in discourse. To gain more insight into the trends in and between advertisements and articles, we applied two techniques from econometrics and complexity science to the (relative) frequencies of words. First, we applied the Granger causality test to determine whether trends in word use in ads impacted trends in articles or vice versa. Second, we analyzed the persistence of words using fractal analysis to identify whether ads differed from articles in terms of dynamics related to word use. Put differently, we determined whether advertising discourse was distinct in its behavior from discourse in articles. Also, we set out to identify specific words that \textit{stuck} with people? The later analysis could point towards the existence of what has been called communicative memory.

\subsection{Communicative Memory}
Communicative memory is one of the many concepts in the field of memory studies. It has been linked to heterogeneous concept cultural memory \cite{donald_mind_2001,assmann_cultural_2011}. Jan Assman has described cultural memory as ``a collective concept for all knowledge that directs behavior and experience in the interactive framework of a society and one that obtains through generations in repeated societal practice and initiation \cite[p. 125]{assmann_collective_1995}.'' One of these repeated societal practices is language use, which also shapes our collective understanding of a shared culture. According to Assman, cultural memory is formed over large periods of time whereas communicative memory represents memories shaped over shorter time spans (80-100 years). In a way, communicative memory can be viewed as the short-term memory of a society \cite{erll_communicative_2008,assmann_collective_1995}.
 
In this paper, communicative memory is approximated by time series of word frequencies newspapers for a period of a century. Memory in a time series is modeled as the presence of self-similarity, or more precisely persistent correlation, between the values of these features at various time steps. By modeling time series of word use in newspapers, we try to capture expressions of communicative memory in public discourse. Merely glancing at visualizations produced by n-gram viewers might show sudden peaks or slow decays in word use. In this paper, however, we quantify such behaviors and show the extent to which word use exhibits particular memory functions.

\section{Methods}

\subsection{Data}
The National Library of the Netherlands (KB) has digitized thousands of Dutch historical newspapers using optical character recognition (OCR) software.\footnote{ These newspapers can be accessed through Delpher: \url{https://www.delpher.nl}} This software turns scans of physical pages into machine-readable data. Unfortunately, the text extracted from the digital scans is often flawed due to imperfections in the original material or limitations of the recognition software. These material blemishes cause the software to not recognize and transcribe every word correctly, which has resulted in conjoined words, complete gibberish, or words in which certain characters were replaced. The age and quality of the original material are important determinants of the software's ability to correctly recognize the text; hence, older newspapers contain many more errors than more recent papers. For this reason, we focused on twentieth-century newspapers. Also, the KB does not provide suitable metrics on the quality of the OCR'ed text \cite{traub_impact_2015}. The study, therefore, assumes that OCR errors are uniformly distributed over the period.

For analysis, we relied on two subsets of the digitized newspaper data. The first subset consisted of the entire set of advertisements ($n_1 = 18,564,411$) in the KB's digitized newspaper archive. The second set held newspaper articles ($n_2 = 11,465,220$) from two national newspapers: \textit{De Tijd} (1890-1974) and \textit{De Telegraaf} 1893-1989. During digitization, the OCR software separated articles from advertisements and stored the document type in the metadata, allowing us to select these two types of documents. Because advertisements made up a smaller portion of the newspapers, we selected the entire set of advertisements to make it more comparable to the corpus of articles in terms of size. Also, we narrowed our focus to two national newspapers because these are more likely to represent wider public discourse than regional newspapers.

We calculated keyword frequencies, more specifically, normalized relative daily term frequency per document for these two subsets. We explicitly looked at 265 words (singular and plural forms collapsed) that denoted consumer products. Based on exploratory data analysis using an n-gram viewer, we selected words that appeared in both advertisements and articles throughout greater periods of time with considerable frequency. Brand names were excluded for two reasons. First, brand names often appeared as part of logos in advertisements, making it more difficult to convert these images to machine-readable text. Second, the techniques used in this paper necessitate the existence of the same words over longer periods of time. More often than not there existed multiple brand names for the same products, which were also not used over longer periods of time.

\subsection{Causal Dependencies}
Several techniques can be used to compare lagged values of time series $X$ with values of a second time series $Y$ to model variation in their correlation coefficient as a function of temporal displacement. The most widely used technique is cross-correlation, which is simply used to detect the variation in the correlation between two time series as a function of lag. The Granger causality test goes beyond mere correlation \cite{granger_investigating_1969} and tests for the existence of causal-like dependencies between temporally disjunctive time series of, for instance, words from two sources. The test, which originated in econometrics, is based on the assumption that causality is more than temporal disjunction, it involves directionality between time series. The relation tested by the Granger causality test is often characterized as predictive causality and represented as $X~Granger~cause~Y$ to distinguish it from more direct causality \cite{sugihara_detecting_2012}. At its core, Granger causality, which is related to correlation, expresses if values of time series $X$ contain information that is uniquely predictive of subsequent values in time series $Y$.

For our study, we used the Granger causality test as follows. To identify a \textit{shaping} relation, we test if variation in a specific word frequency for newspaper discourse ($Y$) at time $t$ is predicted by variation in the frequency for the same word in advertisement discourse ($X$) at earlier time steps $t-1 \ldots t-k$.  We test for $X~Granger~cause~Y$, by comparing the performance of the nested `newspaper discourse only' model:

\begin{equation}
y_{t} = \beta_{0} + \beta_{1}y_{t-1} + \ldots + \beta_{k}y_{t-k} + \epsilon
\end{equation}

with the full `newspaper and advertisement discourses' model:

\begin{equation}
y_{t} = \beta_{0} + \beta_{1}y_{t-1} + \ldots + \beta_{k}y_{t-k} + \alpha_{1}x_{t-1} + \ldots + \alpha_{m}x_{t-m} + \epsilon
\end{equation}

to identify which one does the better job at explaining $y_t$ based on the residuals. The zero-model for the hypothesis then is $H_{0}: \alpha_{i} = 0$ for each $i$ of the element $[1,m]$ with the alternative hypothesis being $H_{1}: \alpha_{i} \neq 0$ for at least one $i$ of the element $[1,m]$. We applied the test bi-directionally such that a shaping relation finds support if we can confirm that `$X~Granger~cause~Y$' and reject that `$Y~Granger~cause~X$' in case of a reflecting relationship (the inverse of shaping). Finally, if both `$X~Granger~cause~Y$' and `$Y~Granger~cause~X$' find support this is viewed as support for a more complex relationship between the two time series.

\subsection{Long-range Dependencies}
In addition to the Granger Causality test, we used fractal analysis to identify if words exhibited long-range dependencies.\footnote{ Long-range dependency is also called persistent behavior or long-memory in time series. The terms will be used interchangeably.} Long-range dependency indicates a rate of decay between two points with increasing time intervals that is slower than exponential decay. Analysis of time-dependent change in complex systems---systems composed of many interacting elements---is an important application of fractal analysis.

Some random processes in complex systems are self-affine, that is, fluctuation patterns at shorter time scales are (statistically) similar to fluctuations at longer time scales. In the case of reading, for instance, fluctuations in reading speed are self-affine across multiple time scales, because both reading fluency and comprehension are affected by elements at short time scales (e.g., words and sentences) and longer times scales (e.g, paragraphs and chapters) \cite{obrien_using_2014}. Such fractal behavior is found in a range of culturally relevant processes related to psychology \cite{chater_scale-invariance_1999}, economy \cite{marchant_scale_2008}, sociology \cite{gao_empirical_2017}, health \cite{eke_fractal_2002}, language \cite{gao_culturomics_2012} and music \cite{voss_1/fnoise_1975}. We argue, therefore, that Fractal analysis has great potential for the study historical trends in cultural expressions. This is particularly the case when we are dealing with `big data' consisting of large sets of mostly unknown parameters \cite{gao_culturomics_2012}.

\begin{figure}[h!]
	\centering
		\includegraphics[width=.75\textwidth]{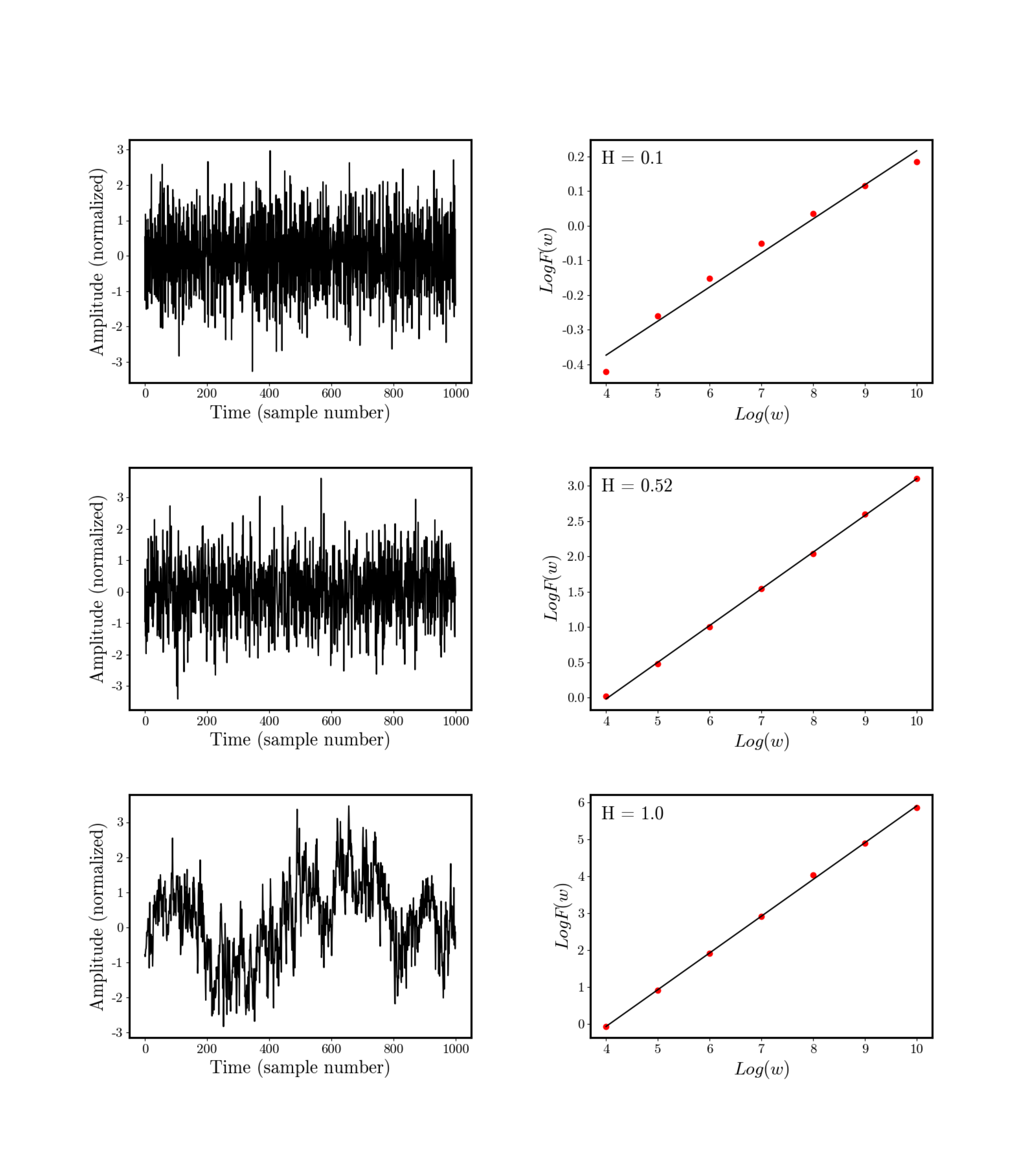}
	\caption{Left: Time series that exhibit anti-persistent (top), short-range (middle), and long-range (bottom) dependencies. Right: Estimation of the Hurst exponent for the matching time series in the left column.}
	\label{fig:hurst}
\end{figure}

We are interested in a particular kind of fractal processes called $1/f^{2H+1}$ processes, in which $H$ refers to the Hurst exponent. The Hurst exponent quantifies the degree of long-range dependencies in a time series, such that when $0 < H < 0.5$, the time series is an anti-persistent process (i.e., a jump up is followed by a jump down, or vice versa, in the increment process), when $H = 0.5$, the time series only has short-range dependencies, and when $0.5 < H < 1$, the time series is characterized by long-range dependencies (i.e. a jump up is followed by another jump up, or vice versa, in the increment process), see figure \ref{fig:hurst}. It is possible the $H > 1$ indicates a non-stationary process. In this study, persistence represents whether a word `stuck' with people and it is in that manner analogous to communicative memory.

\subsubsection{Adaptive Fractal Analysis}
Adaptive fractal analysis (AFA) is a relatively new technique for determining the Hurst exponent of a time series \cite{gao_facilitating_2011,riley_tutorial_2012}. AFA improves the popular detrended fluctuation analysis \cite{peng_mosaic_1994} by identifying a global smoothed trend that can automatically deal with arbitrary, strong nonlinear trends \cite{gao_facilitating_2011}. The technique is based on a nonlinear adaptive multi-scale decomposition algorithm \cite{gao_facilitating_2011}.

After constructing a random walk process from the time series, the initial step of AFA involves partitioning the time series into overlapping segments of length $w = 2n+1$, in which neighboring segments overlap by $n+1$ points. In each segment, the time series is fitted with the best polynomial of order $M$, obtained using standard least-squares regression. The fitted polynomials in overlapped regions are then combined to yield a single global smoothed trend. Denoting the fitted polynomials for the $i-th$ and $(i+1)-th$ segments by $y^{i} (l_1)$ and $y^{(i+1)} (l_2)$, respectively, where $l_1, l_2 = 1,\cdots,2n+1$, we define the fitting for the overlapped region as

\begin{equation}
y^{(c)} (l) = w_1 y^{(i)} (l+n) + w_2 y^{(i+1)} (l),~~l=1,2,\cdots,n+1,
\end{equation}

where $w_1 = \big (1-\frac{l-1}{n} \big )$ and $w_2=\frac{l-1}{n}$ can be written as $(1-d_j/n)$ for $j=1,2$, and where $d_j$ denotes the distances between the point and the centers of $y^{(i)}$ and $y^{(i+1)}$, respectively. Note that the weights decrease linearly with the distance between the point and the center of the segment. Such a weighting is used to ensure symmetry and to effectively eliminate any jumps or discontinuities around the boundaries of neighboring segments. As a result, the global trend is smooth at the non-boundary points and it has the right and left derivatives at the boundary \cite{riley_tutorial_2012}.
 
The global trend can be used to maximally suppress the effect of complex nonlinear trends on the scaling analysis. The parameters of each local fit are determined by maximizing the goodness of fit in each segment. The different polynomials in the overlapped part of each segment are combined such that the global fit will be the smoothest fit of the overall time series. Even if the local fits are linear, $M = 1$, the global trend signal will still be nonlinear. AFA then can be described accordingly: for an arbitrary window size $w$, we determine, for the random walk process $u(i)$, a global trend $v(i), i=1,2,\cdots,N$, where $N$ is the length of the walk. The residual of the fit, $u(i)-v(i)$, characterizes fluctuations around the global trend, and its variance yields the Hurst exponent $H$ according to the following scaling equation:

\begin{equation}
F(w) = \Big [ \frac{1}{N} \sum_{i=1}^N (u(i) - v(i))^2 \Big ]^{1/2} \sim w^{H}.
\end{equation}

By computing the global fits, the residual, and the variance between original random walk process and the fitted trend for each window size $w$, we can plot $\log_2{F(w)}$ as a function of $\log_2{w}$. The presence of fractal scaling amounts to a linear relation in the plot, with the slope of the relation providing an estimate of $H$, see figure \ref{fig:hurst}.

\subsection{Design}
To determine whether advertisements reflected or shaped public discourse, we first applied Granger causality tests to each of the 265 keywords, comparing time series from newspaper and advertisement discourse. We hypothesize the existence of the following three causal-like patterns: 
\begin{enumerate}
	\item Advertisements shaped newspaper articles as expressed by Granger causality directed exclusively from advertisements to articles;
	\item Advertisements reflected newspaper articles as expressed by Granger causality directed exclusively from articles to advertisements; 
	\item A complex, possibly externally-driven, causal pattern as evidenced by cases where Granger causality goes from articles to advertisements \textit{and} vice versa.
\end{enumerate}

For the second step of the analysis, we used AFA to model the persistence for each keyword in both discourses. This enabled us to identify possible dynamic properties of the two types of discourse as a whole and possibly the dynamic properties of particular words. Similar to causal-like patterns, the Hurst exponent has three possible patterns of persistence: anti-persistent processes, short-term correlation processes, and persistent processes. Each keyword's behavior can thus be described by one of nine possible combinations of causality (Granger causality test) and persistence (AFA). Insights into these dynamic properties alongside the causal patterns can help to increase our understanding of the relationship between advertisements and articles, and by extension, between advertisements and society. 

\subsection{Data Analysis}
Statistical tests were conducted with an $\alpha$ level of .005 \cite{benjamin_redefine_2017}. Pearson's correlation coefficient $R$ was used to measure the non-lagged association strength between the time series. We converted Pearson's $R$ using Fisher's Z-transformation to normally distributed z-values to permit averaging. Before applying the Granger causality test for comparison of discourses, lag-1 differencing was used to obtain a stationary keyword time series.

For the analysis of the Hurst exponent for each keyword time series per discourse, we used a simple linear regression and compared this with the constant model. This allowed us to test differences in long-range dependencies between the two different discourses. The Shapiro-Wilk $W$ test confirmed that the distribution of the Hurst exponent did not deviate significantly from normality \cite{shapiro_analysis_1965}.\footnote{ While some keyword time series did show indications of multifractal structure (i.e. local fluctuations with either small or large variation), this information was discarded from the final analysis for the purpose of simplification.}

\section{Results}

\subsection{Directionality}
On average, the variance in correlation for each keyword in all the time series was similar between advertisement and articles. The mean correlation coefficient, $\bar{R}$, between advertisements and the newspapers \textit{De Tijd} and \textit{De Telegraaf} was $\bar{R} =.25$ and $\bar{R} =.27$ respectively. Sixty-two percent (62\%) of these correlations were statistically reliable. The within-discourse correlation, that is, the correlation between \textit{De Tijd} and \textit{De Telegraaf}, was considerably stronger, $\bar{R} =.42$. Seventy-three percent (73\%) of these correlations were significant, suggesting that word use over time in articles between these two newspapers was more similar than between the articles and advertisements. 

\begin{figure}
	\centering
	\includegraphics[width=\textwidth]{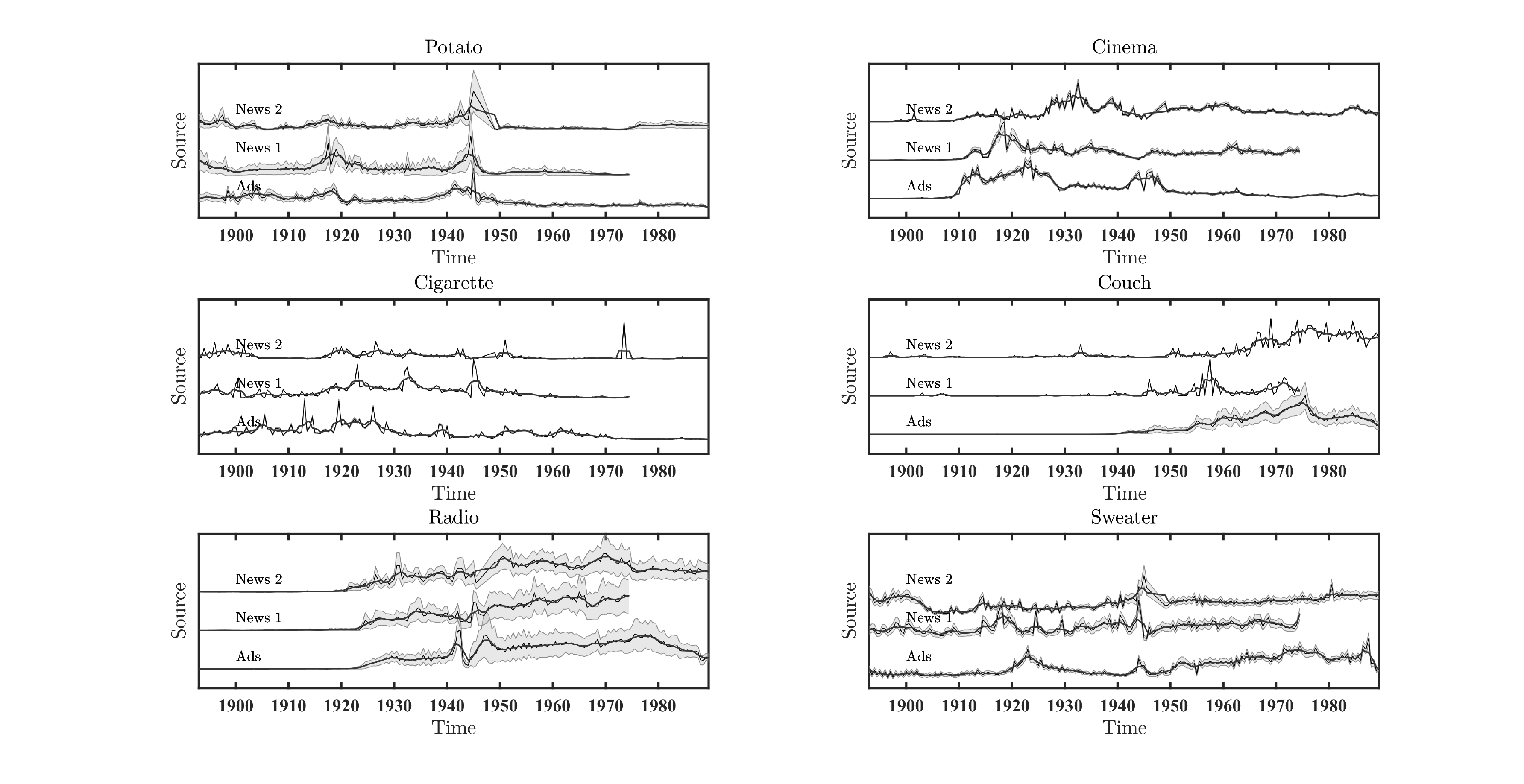}
	\caption{Six keyword frequencies plotted at bi-annual intervals for the two newspapers (News 1: \textit{De Tijd}; News 2: \textit{De Telegraaf}) and the advertisements (Ads). The line is smoothed using a simple moving average filter (window size of five years) and gray bands represent confidence intervals at 95\%}
	\label{fig:signals}
\end{figure}

Analysis showed that there was no overarching causal pattern, but rather multiple causal patterns that were keyword-dependent. Twenty percent (20\%) of the product terms show evidence of a shaping causality, in which discursive trends in advertising discourse uniquely predict those in articles. For 17\% of the product terms, we found the causal pattern in which advertisements reflect articles. Almost half of the product terms (49\%) belong to the complex externally-driven category, that is, both advertisements and articles seem to be primarily impacted by extraneous events that perturb the reflecting-shaping dynamic between the time series. Finally, 14\% of the terms show no indication of predictive causality.

Almost half of the product terms exhibit complex behavior that points towards an external source driving changes in discourse in advertisements and newspapers. These could include economic developments or possibly the invention of new products or technology. In a way, this indicates that the relationship between advertisements and articles was one of negotiation with external developments. Noteworthy keywords in this category were related to produce (`apple', `cauliflower', `lettuce'), energy (`stove', `cokes', `furnace', `gasoline') and audiovisual technology (`tape recorder', `gramophone', `radio', `television').
 
There are slightly more keywords for which advertisements were shaping articles than ads reflecting articles. In case of shaping, we detected that behavior in word use in advertisements was related to behavior in articles. Word that exhibited this behavior referred quite generally to fashion and clothing (`men's clothing', `sweater', `fur', `wool', `flannel', `jeans', `heels'), interior design (`living room', `dining room', `bedroom') and movies (`cinema', `film'). The keywords that exhibited reflecting behavior were more difficult to categorize into particular categories. They ranged from words such as `teapot', `dictionary', to `cheese'. Generally, the words with reflecting behavior seemed to be more specific than the words in the shaping category. For instance, the keywords `chair cushions' and `winter coat' are specific types of cushions and coats.

\subsection{Detection of Communicative Memory}
$H$ for both articles ($t_{1058} = 32.8, p < .00001$) and advertisements ($t_{538} = 38.5, p < .00001$) was significantly higher when compared to a no-memory baseline ($H$: $M = 0.5,~SD = 0.18$). In terms of persistence, discourse in articles and advertisements was different from processes that only showed short-range correlations. AFA found an average difference ($\Delta{H}$) between articles of 0.21. The dynamics related to word use in articles thus differed from advertisements. Word use  in articles was more likely to be a persistent process ($H$: $M = 0.89,~SD = 0.19$) than it was in advertisements ($H$: $M = 1.1,~SD = 0.17$). Notice that advertisements display non-stationary dynamics with $H > 1$. To test whether the difference between ads and articles was significant, we ran linear regression to predict \textit{H} as a function of advertisement and articles (advertisement as baseline). Compared to the constant model, a statistically significant regression model was found ($\chi^{2}_{1} = 149.1, p < .00001$) showing that \textit{H} was, indeed, reliably lower for articles than for advertisements (articles: $\beta$ = -0.18 , SE $\pm$0.01, $F_{1,793}= 163.9$ $p < .00001$). 

On the whole, word use in articles more clearly expressed persistent trends, while word use in advertisements tended to be more irregular, displaying bursts of high activity followed by little or no activity. This indicates that articles more closely express behavior that could be interpreted as communicative memory, while ads seem more haphazard and perhaps catalytic to the construction of communicative memory. This finding resonates with Marchand's claim that advertisements contributed to the ``shaping of a `community of discourse,' an integrative common language shared by an otherwise diverse audience \cite[p.XX]{marchand_advertising_1985}.''

Our findings indicate three distinct types of persistent trends, see table \ref{tab:tab1}. Persistence in only articles, persistence in advertisements and articles, and a lack of persistence in either source. Words that only exhibited persistence in articles included products related to interior design (`living room', `couch', `lamp', `bedroom'). This suggests that discourse about these products was part of a shared language but not clearly part of an advertising discourse. Conversely, words that showed persistence in both discourse included those related to cigarettes but not cigars and tobacco ('cigarettes'), fashion (`fur', `jeans'), energy (`cokes', `furnace', `gasoline') and produce (`apple', `cauliflower', `lettuce'). This suggests that advertising discourse for these products was much more persistent and relied on an established frame of reference. Interestingly, cigarettes and fashion are often presented as typical examples of strongly branded products \cite{blaszczyk_producing_2011,hill_advertising_2002,white_physician_2012,brandt_engineering_2004}. Advertisements for produce, on the other hand, might exhibit persistent processes due to its highly seasonal and reoccurring nature. Keywords that showed no persistence were related to technology (`cinema', `tape recorder', `gramophone', `radio').

\begin{table}
  \newcolumntype{+}{>{\global\let\currentrowstyle\relax}}
  \newcolumntype{^}{>{\currentrowstyle}}
  \newcommand{\rowstyle}[1]{\gdef\currentrowstyle{#1}%
    #1\ignorespaces
  }

  \centering
  \begin{tabular}{+>{\bfseries}l^c^c^c^c}
    \hline
    \rowstyle{\bfseries}
     & Persistence in art. & Persistence in ads \& art. & No persistence\\
     & living room & cigarettes  & cinema \\
     & dining room  & fur  & film \\
     & bedroom & wool & tape recorder \\
     & chair & flannel & gramophone \\
     & couch & jeans & radio \\
     & cupboard & heels & television \\
     & seat & apple & \\
     & lamp & cauliflower & \\
     & & lettuce & \\
     & & cokes & \\
     & & furnace & \\
     & & gasoline & \\
         \hline
  \end{tabular}

  \caption{Example of keywords grouped on type of persistent trend (persistence in articles only, persistence in articles and advertisements, and no persistence}
  \label{tab:tab1}
\end{table}

\section{Discussion}
Using AFA, we found a significant difference in persistent behavior between word use in advertisements and articles. The latter exhibited long-term dependencies whereas advertisements displayed more non-stationary and irregular behavior. In general, advertisements introduced terms, but many of these terms did not persist and their decay was rapid. For articles, on the other hand, keywords that denoted products showed more persistent behavior and were either mentioned recurrently in a self-reinforcing manner or decayed much slower than advertisements. We speculate that this reflects an overarching media dynamic in which ads introduced keywords and articles represented how these products became part of public discourse. However, this dynamic does not hold for all products as evidenced by table \ref{tab:tab1} and, at least partially, by keywords that exhibit a reflecting causal pattern.

The time series of keywords between the two newspapers were more clearly correlated than between the newspapers and advertisements. This shows that word use in newspapers more closely followed each other than word use between advertisements and newspapers. Along with the found different in $H$ between articles and advertisements, the difference in correlation adds evidence to the hypothesis that the dynamics of discourse in ads are different from articles. This suggests that advertisements are not merely a lens on the past, but more clearly a distorted mirror that is shaped to a certain degree by advertisers and its own dynamics.

In terms of directionality, we did not find one dominant pattern. For 20\% percent of the keywords, advertisements reflected articles, and for 17\% of the keywords, advertisements shaped articles. But for almost half of the keywords, there was a more complex causal relationship, indicative of external forces. This lends support to Cowan's argument for a complex interaction pattern in which advertisers, distributors, producers, and consumers negotiated the meaning of a product \cite{cowan_proposal_1997}. 

\subsection{Product Groups}
The causal direction and type of persistence seems to be, to some extent, related to the type of product. We were not able to identify specific categories of keywords in the \textit{reflecting} causal category. However, the complex relationship and \textit{shaping} category offered interesting groupings of words. The groupings made on the basis of the existence of memory and causal directionality leads to the following four points of discussion.

First, products with a shaping dynamic \textit{and} long-term dependencies in articles might point towards products that are not constantly advertised --- expressed as the lack of persistence in ads --- but that nevertheless are part of the cultural life of Dutch consumers throughout the twentieth century, such as bikes, pets, interiors, and clothes. The shaping dynamic reveals that ad makers might have pushed the popularity of these products, which can be described as lifestyle products. One could argue that advertisers might have been able to affect the longevity of these products, effectively installing them within communicative memory.

Second, one of the most noteworthy behaviors is associated with the cigarette. This product exhibits persistence in advertisements \textit{and} in articles, and it shows a shaping causal behavior. This suggests that in advertising discourse and newspaper discourse, cigarettes were a recurring topic that built upon earlier discourse. Moreover, advertisers seemed to be able to shape newspaper discourse on cigarettes. This finding is in line with scholars that view advertisements for cigarettes as the prime example of social engineering \cite{brandt_engineering_2004}. Our study finds that, at least for the Dutch context, cigarette advertisements were a steadily successful form of advertising. The unique behavior of cigarettes was underlined by the fact that related products such as tobacco and cigars behaved dissimilarly. Tobacco and cigars exhibit no persistence and are driven by a complex causal relationship, underscoring different advertising dynamic than found for cigarettes.

Third, some products revealed persistence in both advertisements and articles without displaying a uniform causal relationship. These products include produce, energy sources, and computer systems. One interpretation might be that produce was of prolonged importance (indicated by the existence of long-range dependencies), yet its importance was not driven by advertisers but by an external source. In the case of produce, this external source could be seasonal or economic shifts. The other two product groups (energy sources and computer systems) were also quite instrumental in society, albeit not for prolonged periods during the twentieth century. The keywords associated with energy were most dominant in the first half of the century, whereas, the words associated with computers only appeared in the latter quarter of the century. Nonetheless, they both still exhibited persistence in newspaper articles.

Finally, keywords associated with technological innovations showed two distinct types of behavior. First, keywords such as `cinema', `tape recorder' and `television' did not exhibit any persistence, which could have been the results of the constant innovation and disruptions in the field of audiovisual technology. Another explanation could be the use of different keywords to refer to similar technologies. Further research is needed to explore this behavior related to technology. Second, we found a distinction in the causal relationship between types of technology. `Cinema' and `film' showed a clear causal relationship between ads and articles. The causal relationship might have resulted from the fact that advertisements played an important role in pushing these innovations to a wider audience. Keywords associated with household technology (`radio' and `television'), on the other hand, displayed the complex type of causality. These technological products might be more closely related to particular economic, seasonal, or innovative cycles. Again, further research is needed to untangle these dynamics. 

\section{Conclusion}
Through two data experiments, we have found evidence of a fundamental difference between the dynamic behavior of word use related to consumer products in articles and advertisements published in newspapers. Articles --- taken as a proxy for public discourse --- exhibit persistent trends that are likely to be reflective of communicative memory. Contrary to this, advertisements have a more irregular behavior characterized by short bursts and fast decay, which, in part, mirrors the dynamic through which advertisers introduced terms into public discourse. On the issue of whether advertisements shaped or reflected society, we found particular product types that seemed to be collectively driven by a causality going from advertisements to articles. Generally, we found support for a complex interaction pattern dubbed the consumption junction. Finally, we discovered noteworthy patterns in terms of causality and long-range dependencies for specific product groups.
 
This study shows how methods from fields of econometrics and complexity science can be applied to improve our understanding of complex cultural-historical phenomena. Further research that employs more extensive keyword lists that also includes brand names and cross-cultural comparisons will make it possible to propose a more detailed and general account of the mechanics that underpin this consumption junction.

\subsection{Acknowledgements}
Part of this research was performed while the authors were visiting the Institute for Pure and Applied Mathematics (IPAM), which is supported by the National Science Foundation. The newspaper data was provided by the National Library of the Netherlands (KB).

\bibliographystyle{plainnat}
\bibliography{main}

\end{document}